# Investigating the Dynamics of Social Norm Emergence within Online Communities

Shangde Gao[1], Yan Wang[2*], My T. Thai[3]

1 Ph.D. Candidate, Department of Urban and Regional Planning and Florida Institute for Built Environment Resilience, College of Design, Construction and Planning, University of Florida, 1480 Inner Road, Gainesville, FL, 32601, U.S.; Email: gao.shangde@ufl.edu; ORCID: 0000-0003-2218-2872.

2*Assistant Professor, Department of Urban and Regional Planning and Florida Institute for Built Environment Resilience, University of Florida, P.O. Box 115706, Gainesville, FL 32611, U.S. (*corresponding author*); E-mail: yanw@ufl.edu; ORCID: 0000-0002-3946-9418.

3 Professor, Department of Computer & Information Science & Engineering and Warren B. Nelms Institute for the Connected World, University of Florida, Gainesville, FL 32611, U.S.; E-mail: mythai@cise.ufl.edu; ORCID: 0000-0003-0503-2012.

**Abstract**: Although social norms' effect on mitigating misinformation is identified, scant knowledge exists about the patterns of social norm emergence, such as the patterns and variations of social tipping in online communities with diverse characteristics. Accordingly, this study investigates the features of social tipping in online communities and examines the correlations between the tipping features and characteristics of online communities. Taking "the side effects of COVID-19 vaccination" as the case topic, we first track the patterns of tipping features in 100 online communities, which are detected using Louvain Algorithm from the aggregated communication network on Twitter between May 2020 and April 2021. Then, we use multi-variant linear regression to explore the correlations between tipping features and communities' characteristics. We find that social tipping in online communities can sustain for two to four months and lead to a 50% increase in populations who accept the normative belief in online communities. The regression indicates that the duration of social tipping is positively related to the community populations and original acceptance of social norms, while the correlation between the tipping duration and the degrees among community members is negative. Additionally, the network modularity and original acceptance of social norms have negative relationships with the extent of social tipping, while the users' degree and betweenness centrality can have significant positive relationships with the extent of tipping. Our findings shed light on more precise normative interventions on misinformation in digital environments as it offers preliminary evidence about the timing and mechanism of social norm emergence.

## 1 Introduction

The extensive development of online platforms has fostered the spread of messages generated by stakeholders at various levels, e.g., governmental agencies and individual users, during public events (Y. Wang et al., 2021). A large proportion of user-generated online messages contain inaccurate and misleading information, i.e., misinformation (Del Vicario et al., 2016; Wang et al., 2022). The wide diffusion of misinformation has threatened human society from multiple perspectives, e.g., interfering with collective decision-making on democratic, environmental, and public health issues (West & Bergstrom, 2021). There is an emergent need for suppressing misinformation spreading and mitigating the negative consequences of online misinformation on human society (West & Bergstrom, 2021). Existing studies (e.g., D. T. Nguyen et al. (2012), N. P. Nguyen et al. (2012), Zhang, Alim, et al. (2015, 2016), Zhang et al. (2018), Zhang, Kuhnle, et al. (2016), Zhang, Zhang, et al. (2015)) tend to suppress misinformation with (i) debunking, i.e., correcting the misinformation after people are exposed to it, and (ii) prebunking, i.e., helping people recognize the false/misleading contents (U. K. H. Ecker et al., 2022; Lewandowsky & van der Linden, 2021). The debunking strategy is widely adopted to provide targeted countermeasures for misinformation of specific topics (U. K. H. Ecker et al., 2022), e.g., provide messages with factual elaboration (Gao et al., 2021; van der Meer & Jin, 2020; Wang et al., 2022), fact-checking content (Humprecht, 2020), and messages that stimulate the health-protective measures (Humprecht, 2020). The debunking strategy is not always effective when the explanations that support the misinformation exist widely (Chan et al., 2017). The effect of debunking messages tends to be short-term and washed out by future exposure to misinformation (Mourali & Drake, 2022). Also, the debunking strategy can only be conducted after people's initial exposure to the misinformation (van der Meer & Jin, 2020), while the negative consequences of misinformation may already exist and cause notable social costs.

On the contrary, the prebunking strategy is potentially an effective vehicle that overcomes the limitations of the debunking strategy and confers large-scale resistance against misinformation among the public (van der Linden et al., 2020). The prebunking strategy is based on the social psychological theory of "inoculation". If people are pre-warned and form the belief of rejecting misinformation, they might be "immune" to misinformation (Lewandowsky & van der Linden, 2021). Compared to the debunking strategy, the prebunking strategy focuses on influencing people's beliefs on the topics of misinformation, posing long-term effects on the public and reducing the occurrence of negative consequences of misinformation (Basol et al., 2021). When being implemented at a large scale, the pre-bunking strategy is conducted with *social norm interventions*, which aim to generate the social norms and consensus that support the factual evidence and reject misinformation (Dow et al., 2022).

The basis of social norm interventions is people's adherence to the surrounding social norms (Constantino et al., 2022). Existing in both the digital and physical world (Gao et al., 2022), social norms, i.e., the shared beliefs or acceptable behaviors in communities, have shown a significant relationship with people's belief in the content of misinformation (Andı & Akesson, 2021; Gimpel et al., 2021; Lapinski & Rimal, 2005). Adhering to social norms can satisfy a desire to avoid sanctions, confer benefits by coordinating with others, and provide a simple heuristic about what is accepted/wise in a particular context (Constantino et al., 2022). Based on this psychological phenomenon, social norm interventions have been implemented to help form the belief of supporting factual evidence and rejecting misinformation in both the physical and digital realms (Andı & Akesson, 2021; Gimpel et al., 2021; Lapinski & Rimal, 2005), such as suppressing misinformation about climate actions and health behaviors (Constantino et al., 2022; U. K. Ecker et al., 2022). Specifically, by showing individuals the text that describes the "common beliefs" (i.e., social norms) towards the misinformation of a certain topic, individuals tend to modify their beliefs to match the "common beliefs" and reduce the reliance on the misinformation (U. K. Ecker et al., 2022). In another case, by showing individuals a message that "most responsible people think twice before sharing articles" (a social norm), individuals are not likely to share social media articles that contain misleading or contested content (Andı & Akesson, 2021).

Though the role of the social norm in suppressing misinformation has been identified (Dow et al., 2022; Constantino et al., 2022; U. K. Ecker et al., 2022), scant empirical evidence has been provided to inform the implementation of social norm interventions. Several knowledge gaps and challenges remain. First, with the controlled experiments in physical worlds, recent works have identified that social norm emergence in their artificially designed communities tended to have a tipping process, i.e., social tipping (Berger, 2021; Centola et al., 2018; Ehret et al., 2022). Social tipping is a process that when the "tipping point" is reached, a small change in an individual community can create abrupt, nonlinear change in the acceptance of the normative beliefs across the community (Berger, 2021). By predicting the occurrence and extent of social tipping, policymakers can improve the effectiveness of the social norm interventions by adjusting the timing and efforts of implementing the interventions (Andreoni et al., 2021; Ehret et al., 2022). However, due to the lack of analysis of the online communities, it is unclear whether social tipping also exists in online communities and follows certain patterns regarding the tipping features, e.g., the duration and extent of social tipping. Little knowledge exists to guide the practices of social norm interventions regarding the timing and efforts that are needed to promote the tipping process of norm emergence. Second, experiments in existing studies have identified some evidence regarding the potential relationships between community characteristics and the diffusion of normative beliefs (Hu & Leung, 2017; Savarimuthu & Cranefield, 2011; Sen & Sen, 2010; Yu et al., 2014). However, these experiments were generally based on artificially designed communities in real-world or virtual scenarios, and the experiment findings may not be applicable in the communities of the online environment. Also, how the social tipping process varies in the community characteristics has not been disclosed in the existing studies. There is a need for empirical studies that explore the relationships between community characteristics and social tipping based on real-world communities, providing a reference for the design of social norm interventions.

To fill this research gap, this study aims to answer the following research questions (RQ):

- RQ1: Does social tipping exist during the social norm emergence of online communities? If so, what are the characteristics and patterns of social tipping?
- RQ2: Do the features of social tipping correlate with different network characteristics of individual communities?

This study takes the case of the norms on Twitter regarding the side effects of COVID-19 vaccines. The diffusion of vaccine-related misinformation has led to severe consequences during the pandemic (Loomba et al., 2021). A survey in 2020 showed that more than 55% of U.S. adult participants became hesitant in obtaining COVID-19 vaccines because they believed in the misinformation about the side effects, political issues, and safety issues of the vaccines (Graham et al., 2020). When exposed to misinformation about COVID-19 vaccines, people can become hesitant to take the COVID-19 vaccines, exacerbating their risks to be infected (Loomba et al., 2021). There is an emergent need for suppressing misinformation spreading and mitigating the negative consequences of online misinformation on human society. We utilize Louvain Algorithm (Blondel et al., 2008) to extract the communication communities between Twitter accounts from the tweets containing the topics of COVID-19 vaccines. We adopt the definition of "beliefs" from existing psychological studies (Camina et al., 2021; Durando et al., 2016; Herzog et al., 2013; Ritchie et al., 2021) and focus on if a user thinks the manipulated "side effects" of COVID-19 vaccines exist and accepts/rejects the COVID-19 vaccination. Regarding this case, "supporting COVID-19 vaccination" is our desired online social norm and we investigate the social tipping of the expressed normative belief across communities. We further examine how the dynamics of norm emergences vary across community characteristics, such as modularity and betweenness centrality (Winkelmann et al., 2022). The study contributes to disclosing the temporal patterns and mechanisms of social norm emergence in the online environment. Our findings can facilitate the strategic design of normative interventions for precisely mitigating the dissemination of misinformation in the online environment.

## 2 Data and Methods

### 2.1 Overview

As shown in **Fig. 1**, this study starts by collecting real-time tweets regarding the COVID-19 vaccines and related misinformation using Twitter Streaming API (Twitter, 2022). We define communities in the online environment based on Newman (2003), i.e., groups of vertices that have a high density of edges within them, with a lower density of edges between other groups. Specifically for this study, we detect communities from the "retweeting" and "mentioning" networks among Twitter users in the whole study period. For example, if one Twitter user retweets/mentions another user within the whole study period, one edge will exist between these two users. Among the identified individual communities, we select those with a relatively large population (i.e., more than ten users) and long periods of existence (i.e., more than ten days). With these communities, we track the temporal change of the community population that follows the normative belief (i.e., tracking norm emergence) and extract the community characteristics (e.g., modularity, average degree). After preparation, we first answer RQ1 by observing if social tipping can be identified in the temporal trend of social norm emergence in our detected individual communities. If tipping exists, we capture the patterns of the features of social tipping, which include the tipping extent and duration in this study. Based on the tipping features and community characteristics, we answer RQ2 and explore if significant correlations exist between social tipping and community characteristics.

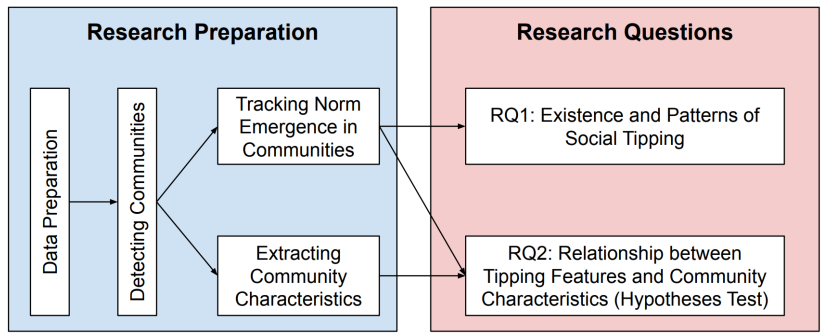

Figure 1 Research procedure

*2.2 Data Preparation*

The basic dataset is collected with Twitter Streaming API between May 1, 2020, and April 30, 2021, regarding COVID-19 vaccines. Specifically, we use keywords of COVID-19 vaccinations to filter out the tweets that are related to COVID-19 vaccines, including the keywords of "vaccine," "vax," "vaccination," and brands of COVID-19 vaccines, e.g., "Pfizer". We extract the online communities based on the communication networks such as "mentioning/replying" messages (i.e., "@username") and retweeting messages (i.e., "RT @username") for multiple reasons. First, retweeting/replying behaviors tend to happen between the users who have following relationships and represent the active social ties between online users (Ozer et al., 2016; B. Wang et al., 2021; Weitzeil et al., 2012). Especially, a study of retweets about COVID-19 (B. Wang et al., 2021) indicated that more than 50% of the retweets about COVID-19 information were generated between users with follower/following relationships. Second, retweeting/replying behaviors can well reflect the social influence of social media users, as the users who tend to retweet or reply to the messages from others if they are influenced by the tweet content (Evkoski et al., 2021; Yuan and Crooks, 2018). We can potentially capture how a certain belief diffuses among social media users based on the interactions between the users (e.g., retweeting/replying to tweets) (Evkoski et al., 2021).

Based on the summary of COVID-19 vaccine-related misinformation from Skafle et al. (2022), we focus on the "side effect" topic of COVID-19 vaccines from the collected tweets, which generally discuss: (a) whether COVID-19 vaccines have side effects that can heavily threaten human health, (b) whether COVID-19 vaccines can make people killed, and (c) whether COVID-19 vaccines have not passed trials and are poisonous. We use keywords (**Table 1**) of these three topics to identify the related tweets in our collected dataset. The keywords in the pattern of "word A + word B", represent the queries that a tweet is regarded as relevant to the topics if both "word A" and "word B" can be identified in the main text of the tweet. The nodes in the online individual communities are the users of the tweets in the basic dataset. We only keep the users whose tweets mentioned other users in the basic dataset, or the users who have been mentioned by other users in the basic dataset. The news bot accounts are also removed. We finally extract 19,839,188 tweets containing the keywords about the three topics of misinformation that were posted by 5,462,900 distinct users (see **Table 1**). We furtherly detect individual communities and analyze the norm emergence with this dataset.

Table 1 Keywords of misinformation related to the side effects of COVID-19 vaccines

| Topics | Keywords |
| --- | --- |
| COVID-19 vaccines have side effects | "side effect", "autism", "autistic", "mental+illness", "psychological+illness", "mental issue", "psychological issue", "infertility", |
| COVID-19 vaccines can make people killed | "children+die", "children+died", "children+dying", "soldier+die", "soldier+died", "soldier+dying", "old+die", " old +died", " old+dying", |
| COVID-19 vaccines have not passed trials and are poisonous | "skip+trail", "poison", "not tested", "doesn't be tested", "isn't tested", "aren't tested", "didn't be tested", "wasn't tested", "weren't tested", "haven't been tested" |

## *2.3 Community Detection*

In the retrieved communication network, the edges between users are formed when users reply to or retweet from other users. The weights of the edges are the frequencies of one user mentioning the other user within one day. We detect individual communities from social networks using Louvain Algorithms (Blondel et al., 2008). Louvain Algorithm is a combinational optimization algorithm that aims to maximize the modularity among the detected individual communities. The algorithm has a process that first assigns every node to be in its community and then for each node it tries to find the maximum positive modularity gain by moving each node to all its neighbor communities. If no positive gain is achieved the node remains in its original community (Blondel et al., 2008). Compared to other algorithms, Louvain Algorithm can efficiently capture the individual communities from a large-scale network, such as a social media network with millions of users. To better reveal the social tipping in large communities instead of small groups (e.g., a small group with less than ten members), we select the 100 communities with the largest populations among our detected communities for the following analysis.

## *2.4 Classifying Individual Users' Expressed beliefs towards Misinformation about COVID-19 Vaccines and Tracking Norm Emergence in Communities*

Based on the user's tweets, we classify the expressed beliefs of individuals at a certain period regarding the side effect of COVID-19 vaccines. We first classify the expressed beliefs in the tweets of individual users. We train a Long Short-Term Memory (LSTM) model with 2,000 tweets related to COVID-19 vaccination and use this model to estimate if tweets from specific users with expressed beliefs that support or reject misinformation about the side effects of the COVID-19 vaccination. LSTM has a good performance in existing studies regarding text classification because it captures phrase-level and sentence-level feature patterns in the tweet text (Zhou et al., 2018). The validated accuracy and loss of the LSTM classifier during training are shown in **Fig. 2**, which reach 0.8892 and 0.2292 separately after training, and the RMSE of the classification outcomes are 0.3719. These metrics indicate that our LSTM classifier has an acceptable performance in classifying the expressed beliefs of individual users.

After classifying the expressed beliefs delivered in the tweets, we obtain the overall expressed belief of each user on each day based on their tweets on that day. Specifically, we calculate the proportion of tweets that one user generates in one day that rejects the misinformation about COVID-19 vaccines. Specifically, if more than 50% of the tweets are supporting the COVID-19 vaccination, we regard the user accept the COVID-19 vaccination on that day. If only one tweet is generated by one user on one day, we regard the expressed belief in the tweet as the expressed belief of that user on a specific date.

We then aggregate the individuals' expressed beliefs to the community level and track the norm emergence in our sample communities. We regard the normative belief as "rejecting the misinformation about COVID-19 vaccines regarding side effects", and the emergence of norms within a community is tracked by the temporal trend of the proportion of community members who hold the normative belief. From the temporal trends, we may identify the tipping points where the acceptance increased rapidly.

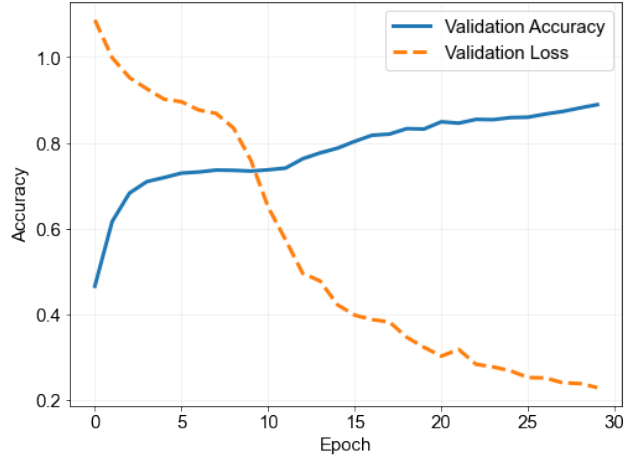

**Figure 2** The accuracy and loss of the LSTM classifier during training

*2.5 Characterizing the Emergence of Social Norm*

Based on the temporal trends of norm emergence in the sample communities, we first observe the trends and detect if social tipping exists in the communities (RQ1). We detect the existence of social tipping according to the tipping's definition, i.e., the increase of community members adopting the norms in specific periods is relatively more rapid than in the past periods (Berger, 2021). We calculate the daily increase in the proportion of community members adopting the normative belief, observing if the increase in a certain period is relatively more rapid than the previous periods. If so, we will regard the social tipping as existing during the norm emergence of our sample communities. If social tipping does exist in the sample communities, we adopt the measurements of social tipping in existing studies (Andrighetto & Vriens, 2022), including the *duration* and the *extent* of the social tipping (illustrated in **Fig. 3**). The duration represents the number of time steps that the social tipping exists. The extent of social tipping is measured as the change in the proportion of community members adopting the normative belief before and after social tipping.

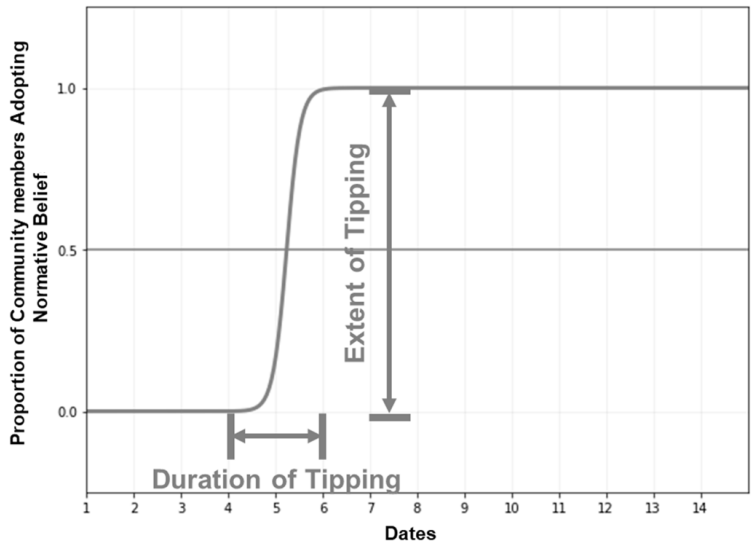

**Figure 3** Illustration of duration and extent of social tipping

*2.6 Investigating Relationships Between Community Characteristics and Tipping Features*

Some characteristics of online individual communities (**Table 2**) may influence the duration of social tipping by increasing or decreasing the rapidness of the tipping process. Some community characteristics may also influence the extent of social tipping by causing large-scale social norm acceptance within the individual community, e.g., the modularity of online individual communities. This study investigates the statistical correlations between the characteristics of online communities (**Table 2**) and the duration and extent of social tipping regarding the proportion of community members accepting the social norms (RQ2).

**Table 2** Characteristics of online communities

| Characteristics | Reference |
| --- | --- |
| Modularity | (Winkelmann et al., 2022) |
| Messaging frequency | (Centola et al., 2018) |
| Network size | (Sabarwal & Higgins, 2021) |
| Original acceptance levels of social norms | (Berger, 2021) |
| Degree and betweenness centrality of community members | (Winkelmann et al., 2022) |

We specify the influence of each community's characteristics on the duration and extent of social tipping when examining each hypothesis. We specifically test the following hypotheses that are designed for each of the community characteristics in **Table 2**.

- *H1: The modularity of a community has a positive relationship with the duration and extent of social tipping.*
- *H2: The average messaging frequency among members in an online community has a positive relationship with the duration and extent of social tipping.*
- *H3: The size, i.e., the number of members, of a community has a negative relationship with the duration and extent of social tipping.*
- *H4: The original proportion of community members who accept the normative belief has a negative relationship with the duration and extent of social tipping.*
- *H5.1: The average degree of network communities has a positive relationship with the duration and extent of social tipping*
- *H5.2: The average betweenness centrality of network communities has a positive relationship with the duration and extent of social tipping*

Before hypothesis testing, we check the statistical distributions of all the considered community characteristics and the features of social tipping. In this way, we can identify if the data of hypothesis testing has an obvious bias. As shown in **Fig. 4**, most communities have modularity that is lower than 0.1. The network size of most communities is smaller than 200 users, and the messaging frequency among the community users tends to be lower than 10 messages a day. For the original acceptance of social norms, most communities have an acceptance level of lower than 40% when the communities emerge. But still, more than twenty communities have the original acceptance that is higher than 80% when the communities emerge. Additionally, the average degree and betweenness centrality of communities tend to evenly distribute in a small range, e.g., 1.8 to 2.0 for the average degree, and 0 to 0.12 for the betweenness centrality.

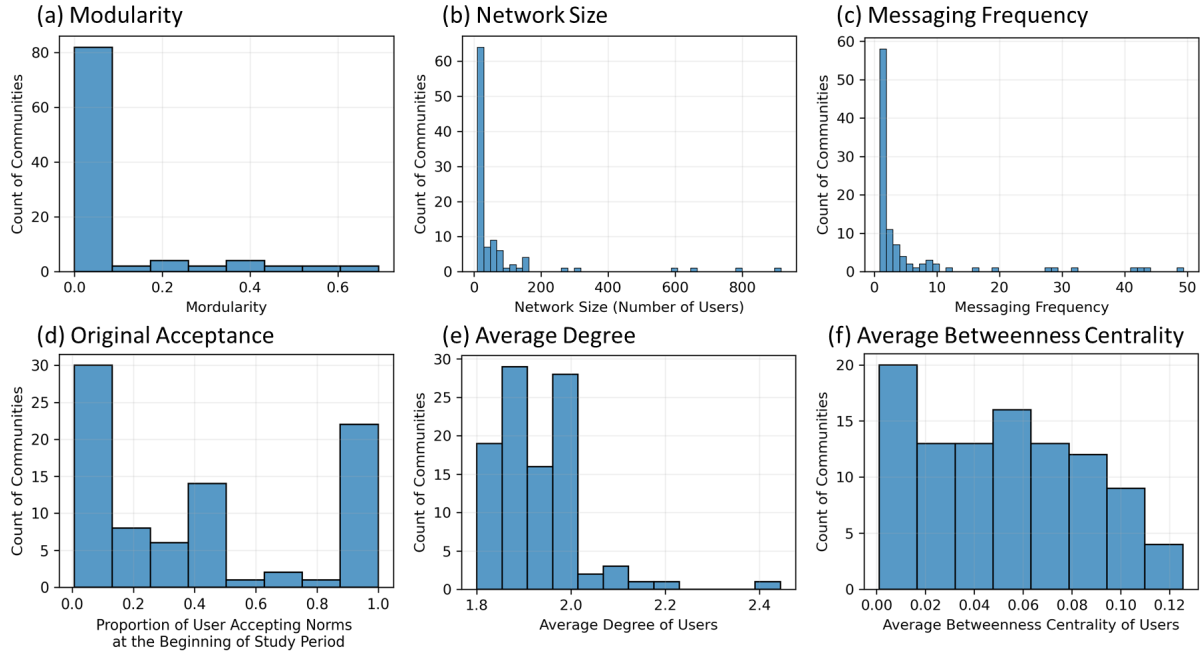

**Figure 4** Distributions of community characteristics

We examine all the hypotheses mentioned above with multi-variant linear regression (Eq. 1 and 2). Based on the identified communities, we examine our proposed hypotheses based on the statistical significance (i.e., $p-value$) and whether the coefficients of community characteristics are positive or negative. For example, to examine hypothesis H1 regarding the duration of social tipping, if the $p-value$ for the variable $Modularity$ is low and the coefficient for this variable is positive, we can state that modularity has a significantly positive relationship with the duration of social tipping in an online community.

$$\text{Duration} \sim Modularity + Message\ Frequency + Netwrk\ Size + Original\ Acceptance \\ + Average\ Degree + Average\ Betweenness\ Centrality \quad (1)$$

$$\text{Extent} \sim Modularity + Message\ Frequency + Netwrk\ Size + Original\ Acceptance \\ + Average\ Degree + Average\ Betweenness\ Centrality \quad (2)$$

## 3 Results

*3.1 Trends and Patterns of Social Norm Emergence in the Sample Communities*

To answer RQ1, we first check the temporal trends of social norm emergence, i.e., the change of norm acceptance among sample communities, aiming to identify if "social tipping" can be identified. Specifically, we determine that social tipping happened within a certain period (e.g., between two specific dates) if the daily change of the proportion of the population who adopt the normative belief (i.e., rejecting misinformation) in the community is much higher than in the past periods. From the temporal trends of social norm emergence in the largest ten sample communities (**Fig. 5**), we find that social tipping does exist, and the social tipping of different communities occurred nearly spontaneously between December 2020 (when the U.S. FDA first issued emergency usage of COVID-19 vaccines (HHS, 2022)) to April 2021. Especially at the end of December 2020, the daily increase of the population who adopt the norms exceeded 10%, which was much higher than the past daily increase (which tended to be lower than 4%). After tipping in these communities, the populations that hold the normative belief towards COVID-19 vaccination in each community generally reached 65% after three months of social tipping.

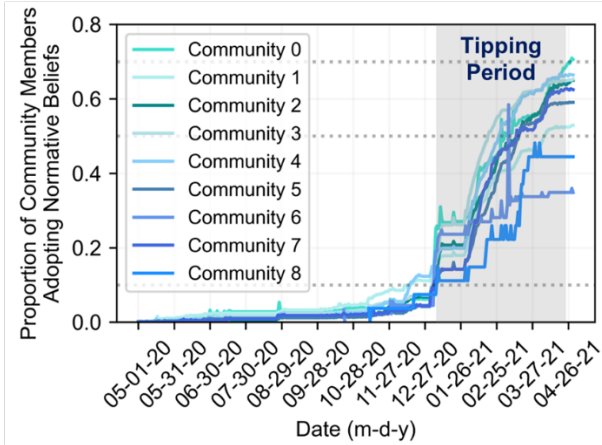
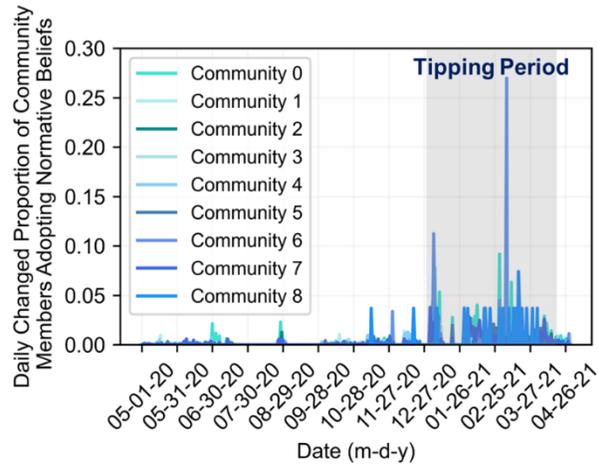

**Figure 5** Temporal trends (a) and daily change (b) of social norm emergence in the ten largest sample communities

Based on the social tipping we identified, we further check the statistical distributions of features of social tipping among our detected communities, shown in **Fig. 6**. The histograms of the tipping durations and extent indicate that, for norms related to the misinformation about COVID-19 vaccines, the social tipping in online communities tends to be relatively long-term and intense. Specifically, the average duration of tipping is 83.26 days, the median tipping duration is 96.5 days, and 95% of the sample communities have a tipping duration between 59 days and 103 days. 17% of the communities have durations that are shorter than one month, and the duration of 8% of the community is longer than four months. For the extent of tipping, we can identify that the increase of population who adopt the normative belief in 86% of the sample communities exceeds 40%, and the tipping in 56% of the sample communities even has the extent that exceeds 50%. Overall, social tipping in online communities regarding the norms of rejecting misinformation tends to exist for two to four months, and the tipping extent in more than half of the online communities may exceed 50%.

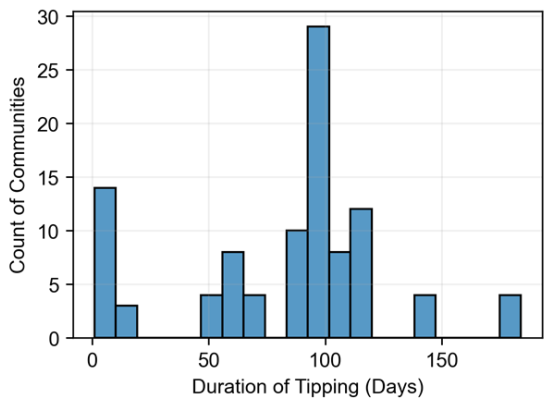
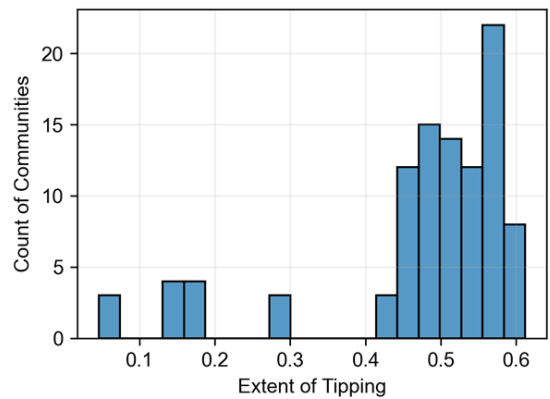

**Figure 6** Distributions of features of social tipping among detected communities

*3.2 Relationships Between Community Characteristics and Tipping Features*

Before conducting the regression, we first check the dependence of the community characteristics, and the outcome is shown in **Fig. 7**. Specifically, the absolute value of the correlation between each pair of community characteristics is lower than 0.2. The test outcomes indicate that the considered community characteristics in this study are relatively independent of other characteristics and can be included in the multi-variant linear regression models.

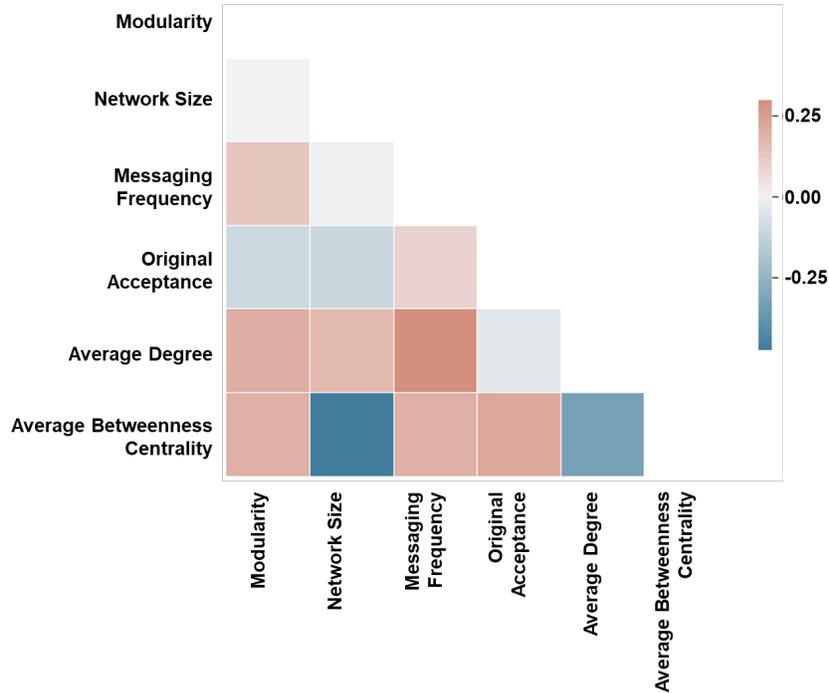

**Figure 7** Dependence test of community characteristics (whiter colors represent lower correlations)

*Relationships between community characteristics and the duration of social tipping.* The table of the regression outcomes is shown in **Table 3**, and the estimated coefficients and 95% confidence intervals (CI) for each community's characteristics are shown in **Fig. 8**. Among our selected characteristics of the detected communities, the network sizes (H3), original acceptance levels of social norms (H4), and the average degree (H5.1) among users have significantly positive impacts on the duration of social tipping. Specifically, although not significant, modularity and communication frequency among community members have a negative relationship with the duration of social tipping (H1, H2). The high-level betweenness centrality in online communities has a positive relationship with the duration of social tipping, but this relationship is not significant (H5b). Based on the outcomes of this regression outcomes, we identify that social tipping is highly related to the context and interactions among the community members. Specifically, the high-level average degree indicates that each community member can communicate with a large number of peers within the community. The original proportion of community members adopting the normative belief indicates the context literacy of the community members regarding the topics of misinformation. Our results indicate that social norms can spread more easily if the individuals are exposed to the information and interact with more peers than the communities with few interactions. Also, the community members who originally do not reject the misinformation may not easily change their belief if they can expose to many interactions with their peers that originally reject the misinformation (i.e., high-level original acceptance). Additionally, the speed of norm emergence may not increase in the large-scale communities, making the duration of tipping longer in the large-scale communities than in the small communities.

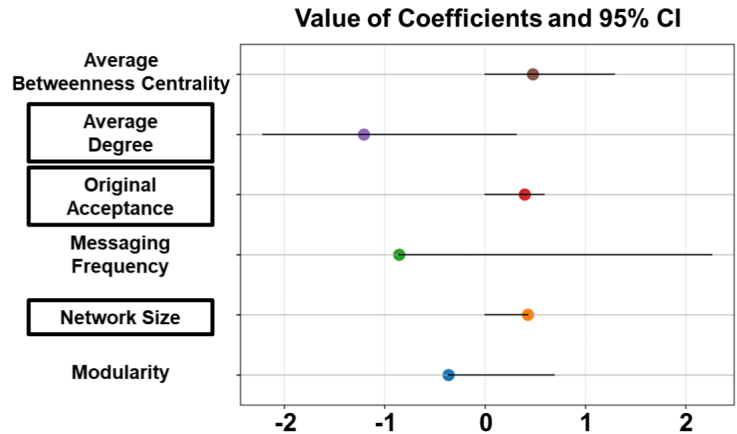

**Figure 8** Estimated values and 95% CI of coefficients in regression for the duration of social tipping (significant variables are within red boxes)

**Table 3** Outcomes of multi-variant linear regression for the *duration* of social tipping (Adjusted $R^2$: 0.648)

| Variables | Coefficient | Standard Error | t Value | P>|t| | [0.025 | 0.975] |
|---|---|---|---|---|---|---|
| Modularity | -0.3648 | 0.427 | -0.855 | 0.394 | -1.207 | 0.477 |
| **Network Size** | 0.0012 | 0.001 | 2.212 | 0.028* | 0 | 0.002 |
| Messaging Frequency | -0.0011 | 0.002 | -0.545 | 0.587 | -0.005 | 0.003 |
| **Original Accept Level** | 0.6879 | 0.305 | 2.258 | 0.025* | 0.087 | 1.289 |
| **Average Degree of Users** | 0.4688 | 0.08 | 5.879 | < 0.001*** | 0.311 | 0.626 |
| Average Betweenness Centrality of Users | 2.3743 | 2.324 | 1.022 | 0.308 | -2.212 | 6.961 |
| Significance Levels: 0 '***' 0.001 '**' 0.01 '*' 0.05 '.' 0.1 ' ' 1 | | | | | | |

*Relationships between community characteristics and the extent of social tipping.* The table of the regression outcomes is shown in **Table 4**, and estimated coefficients and 95% confidence intervals (95% CI) for each variable of community characteristics are shown in **Fig. 9**. Among our selected characteristics of the detected communities, the average degree (H5a) and betweenness centrality (H5b) among users have a significantly positive impact on the duration of social tipping. Meanwhile, the modularity (H1) and original acceptance (H4) of social norms can have a significantly negative relationship with the extent of social tipping. Different from the average degree and betweenness centrality, the significance of the relationships between other community characteristics and the extent of social tipping is not high. Specifically, network size and communication frequency among community members have an insignificant relationship with the extent of social tipping (H2, H3). Based on the regression outcomes, we identify that the extent of social tipping is also highly related to the context and interactions among the community members. Both the betweenness centrality and degree are related to how closely the community members are connected, and the original acceptance of the normative belief is related to the literacy of community members regarding the topics of misinformation. The positive and high-level influence of average degree and betweenness centrality on the tipping extent indicates that more community members will finally turn to the normative belief if they are exposed to heavy interactions with other community peers. Also, similar to the regression outcomes of tipping duration, the community members who originally do not reject the misinformation may not easily change their expressed belief if they can expose to many interactions with the peers that originally reject the misinformation (i.e., high-level original acceptance).

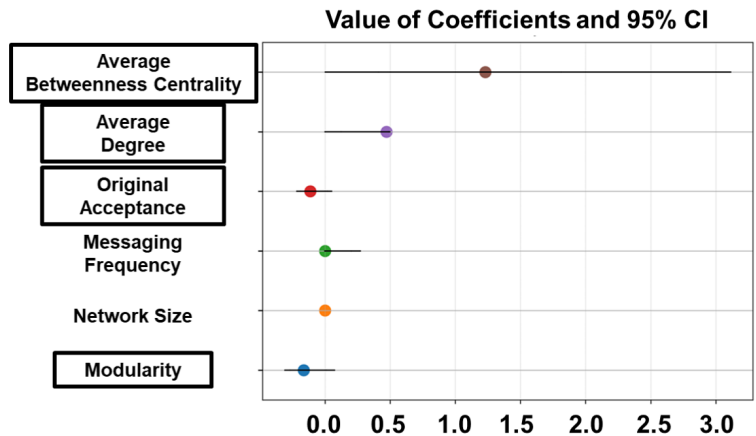

**Figure 9** Estimated values and 95% CI of coefficients in regression for the extent of social tipping (significant variables are within red boxes)

**Table 4** Outcomes of multi-variant linear regression for the *extent* of social tipping (Adjusted $R^2$: 0.972)

| Variables | Coefficient | Standard Error | t Value | P>|t| | [0.025 | 0.975] |
|---|---|---|---|---|---|---|
| **Modularity** | -0.1635 | 0.073 | -2.256 | 0.025* | -0.307 | -0.02 |
| Network Size | -0.0001 | 0.0000897 | -1.271 | 0.205 | 0 | 6E-05 |
| Messaging Frequency | 0.0001 | 0.000 | 0.272 | 0.786 | -0.001 | 0.001 |
| **Original Accept Level** | -0.1127 | 0.052 | -2.176 | 0.031* | -0.215 | -0.01 |
| **Average Degree of Users** | 0.4725 | 0.014 | 34.87 | < 0.001*** | 0.446 | 0.499 |
| **Average Betweenness Centrality of Users** | 1.2297 | 0.395 | 3.113 | 0.002** | 0.45 | 2.009 |

Significance Levels: 0 '***' 0.001 '**' 0.01 '*' 0.05 '.' 0.1 ' ' 1

## 4 Discussion

Social norm interventions can potentially mitigate the spread of misinformation, while insufficient knowledge exists regarding the existence and patterns of social tipping in the online environment, as well as how the tipping features vary in communities with different network characteristics. This study investigates the existence of social tipping in the emergence process of the norms and focuses on rejecting the misinformation about COVID-19 vaccines' side effects. Also, our regression outcomes indicate that the duration of tipping is more correlated to the size, average degree, and original acceptance of the normative belief among the community members. The extent of social tipping (i.e., the increase of community members adopting the normative belief) is more related to the average degree, average betweenness centrality, modularity, and the original acceptance of the normative belief among the community members.

This study advances existing knowledge bodies from several perspectives. First, existing studies focused more on the physical world or artificially designed communities (Berger, 2021; Centola et al., 2018; Ehret et al., 2022), lacking exploration of the existence and patterns of social tipping in the online digital environment. As there can be a difference between the social norm emergence in digital and other environments, existing knowledge of social tipping may not be fully applicable to the online social norm intervention. To fill this gap, we conduct empirical studies with the online communication dataset from Twitter and investigate the social norm emergence in 100 sample communities. To a certain extent, our study helps identify the statistical distributions of the duration and extent of social tipping in online communities. The large datasets and sample communities with various characteristics in this study make it possible to disclose the general patterns of social tipping in online environments.

Second, the existing knowledge body (e.g., Hu & Leung 2017, Savarimuthu & Cranefield 2011, and Sen & Sen 2010) rarely analyzed the relationships between the patterns of social tipping and the network characteristics of online communities, e.g., the modularity of the communities or the degree of community members. To fill this gap, our hypothesis testing with 100 sample communities can help identify the characteristics of online communities that are significantly correlated with the tipping duration and extent. We highlight the significant correlation between the features of social tipping and the modularity, community size, average degree, average betweenness centrality, and original acceptance of the normative belief. Our findings can contribute to disclosing the general relationships between social tipping and community characteristics, supporting the future designing of online social norm intervention strategies.

Limitations still exist in this study and open opportunities for our further studies. First, there are still some external factors that can influence the individuals' expressed belief (changes), such as governmental policies, while this study does not include these factors. Our future studies will include the factors of physical communities to capture the relationships more accurately between social tipping and online community characteristics. Second, this study focuses on the topics of COVID-19 vaccine-related misinformation, of which the community characteristics and social tipping may follow distinct temporal patterns than other topics. To generate more generalizable findings regarding social tipping in online communities, our future studies will study multiple topics of online communications, e.g., other prevention measures for COVID-19. Third, this study regards each individual community as relatively isolated from its neighboring communities, while it is possible that the norm emergence in the neighboring communities also contributes to the social tipping of the individual communities. Our future studies will investigate the norm emergence and social tipping in the circumstance of multi-community social networks and explore the relationships between social tipping in different communities. Fourth, we regard the characteristics of communities as relatively stable in the study period, while community characteristics may be temporally dynamic and have different levels of influence on the norm emergence over different periods. Our future studies will capture the dynamics of online communities and investigate the temporal interactions between the network characteristics of individual communities and the trend of norm emergence.

## 5 Conclusion

Exploring the patterns of social tipping and the relationship between social tipping and community characteristics is critical for tailoring social norm interventions for mitigating online misinformation. Our study contributes to the knowledge regarding the heterogeneous temporal patterns and mechanisms of social tipping in online communities. Our findings can guide public health authorities, emergency responders, and other crisis managers regarding suppressing online misinformation, such as actively disseminating and endorsing messages delivering benign normative beliefs on online platforms. With tailored intervention strategies, crisis managers can motivate the online populations to conduct appropriate prevention measures (e.g., taking COVID-19 vaccines) as well as mitigate the adverse impacts caused by ineffective prevention behaviors (e.g., rejecting vaccinations arbitrarily). With the probunking interventions with social norms, individuals can potentially form positive attitudes towards the public health campaign and proactively reject and suppress the spread of online misinformation.